\documentclass[aps,prd,floatfix,nofootinbib,showpacs,superscriptaddress,twocolumn]{revtex4-2}
\usepackage{color}
\usepackage{graphicx}
\usepackage{dcolumn}
\usepackage{bm}
\usepackage[utf8]{inputenc}
\usepackage{gensymb}
\usepackage{latexsym}
\usepackage{latexsym}
\usepackage{amssymb}
\usepackage{amsmath}
\usepackage{amsfonts}
\usepackage{nicefrac}
\usepackage{float}
\usepackage{booktabs}
\usepackage{siunitx}
\usepackage{slashed}
\usepackage{xcolor}
\usepackage[mathscr,scaled=1.15]{urwchancal}
\DeclareFontFamily{OT1}{pzc}{}
\DeclareFontShape{OT1}{pzc}{m}{it}%
{<-> s * [1.15] pzcmi7t}{}
\DeclareMathAlphabet{\mathpzc}{OT1}{pzc}{m}{it}
\newcommand{\flaligne}[1]{\begin{flalign} #1 \end{flalign}}
\newcommand{\bracket}[1]{\left( #1 \right)}

\begin{document}


\title{The time-like electromagnetic kaon form factor}

\author{A.S. Miramontes}
 \email{angel.miramontes@umich.mx}
\affiliation{%
Instituto de F\'isica y Matem\'aticas, Universidad Michoacana de San Nicol\'as de Hidalgo, Morelia, Michoac\'an 58040, Mexico 
}%

\author{Adnan Bashir}
 \email{adnan.bashir@umich.mx}
\affiliation{%
Instituto de F\'isica y Matem\'aticas, Universidad Michoacana de San Nicol\'as de Hidalgo, Morelia, Michoac\'an 58040, Mexico 
}%

\date{\today}

\begin{abstract}
We compute the electromagnetic charged kaon form factor in the time-like region by employing a
Poincaré covariant formalism of the Bethe-Salpeter equation to study quark-antiquark bound states in conjunction with the 
Schwinger-Dyson equation for the quark propagator. Following a recent kindred calculation of the time-like electromagnetic pion form factor, we include the most relevant intermediate composite particles permitted by their quantum numbers in the interaction kernel to allow for a decay mechanism for the resonances involved. This term augments the usual gluon mediated interaction between quarks.  
For a sufficiently low energy time-like probing photon, the electromagnetic form factor is saturated by the $\rho(770)$ and $\phi(1020)$ resonances. We assume $SU(2)$ isospin symmetry throughout.
Our results for the absolute value squared of the electromagnetic form factor agree qualitatively rather  well and quantitatively moderately so with available experimental data.

\end{abstract}

\maketitle


\section{Introduction}
The light pseudoscalar mesons, pion and kaons, play a crucial role in the description of low-energy dynamics of the strong interactions where emergent phenomena like confinement and dynamical chiral symmetry breaking (DCSB) become dominant. Analyzing the internal structure of these Goldstone modes associated with DCSB is of foremost importance to obtain a better understanding of non-perturbative QCD as well as formation and properties of hadrons. Extensive experimental measurements have been reported on the pion electromagnetic form factor for space-like as well as time-like photons~\cite{CMD-2:2006gxt, CMD-2:2005mvb,Barkov:1985ac, DM2:1988xqd,Brown:1973wr, Bebek:1974iz,JeffersonLabFpi-2:2006ysh,Bebek:1974ww}. In the latter case, it is dominated by the $\rho(770)$ resonance. In a conspicuous contrast, the existing data on the kaon electromagnetic form factor is scarce. For  space-like photons, only a few data points have been reported at low $Q^2 > 0$~\cite{Amendolia:1986ui,Dally:1980dj}, $Q$ being the four-momentum of the probing photon. For time-like photons, existing measurements lie solely in the close vicinity of the $\phi$ resonance~\cite{Achasov:2000am, Akhmetshin:1995vz,Kozyrev:2017agm} whereas the region $-1 < Q^2 < 0$ remains hitherto unexplored.  

Nevertheless, the Electron Ion Collider (EIC) plans to generate a large amount of high precision data on the kaon electromagnetic form factor both for  space-like and time-like photon momenta~\cite{Arrington:2021biu,Aguilar:2019teb}. Additionally, data in the space-like domain is currently being analysed at the Jefferson Laboratory 12 GeV (upgraded) Experiment E12-09-011~\cite{Horn:2010}. Moreover, with increasing effort and precision in experimental measurements of time-like form factors in the upcoming PANDA experiment~\cite{PANDA:2021ozp}, understanding of the QCD dynamics in this intriguing domain is of paramount interest. 

On theoretical side, the space-like electromagnetic
pion form factor has been explored in lattice QCD~\cite{Shultz:2015pfa,Alexandrou:2021ztx,Gao:2021xsm}, a combined analysis of Schwinger-Dyson (SDE) and Bethe-Salpeter equations (BSE)~\cite{Maris:2000sk,Chang:2013nia}, holographic QCD~\cite{Kwee:2007dd} and front form dynamics~\cite{deMelo:2003uk} among other models. The functional approach of SDE and BSE has only recently been extended to evaluate the time-like pion form factor~\cite{Miramontes:2021xgn,Sauli:2022ild}.
Similarly, the kaon form factor has also been examined within different frameworks, for example, lattice QCD \cite{Alexandrou:2021ztx}, chiral perturbation theory~\cite{Bijnens:2002hp}, holografic QCD~\cite{Abidin:2019xwu}, SDE and BSE~\cite{Burden:1995ve,Maris:2000sk}, etc. For the time-like kaon form factor, only limited research is available, for example in dispersion theory~\cite{Stamen:2022uqh} and vector meson dominance models~\cite{Bruch:2004py}.

We study the time-like electromagnetic kaon form factor using a coupled formalism of SDE and BSE which captures the non-perturbative character of QCD excellently well and has produced plenty of
hadron physics predictions. In particular, it has been successfully applied to describe  electromagnetic, axial and transition form factors~\cite{Nicmorus:2010sd,Eichmann:2011pv,Sanchis-Alepuz:2017mir,Chang:2013nia,Raya:2015gva, Raya:2016yuj,Eichmann:2017wil,Ding:2018xwy,Miramontes:2021xgn}. These form factors probe the implications of confinement and DCSB in the infrared and help us understand how a transition to the perturbative realm of QCD takes place by matching onto the predictions of asymptotic QCD within the grasp of the same experiment. 
These form factors also play a pivotal role in testing the very limits of the celebrated standard model of particle physics put forward by Glashow, Weinberg and Salam in precision measurements and calculations of physical observables such as the anomalous magnetic moment of muon,  see for example~\cite{Eichmann:2019tjk,Raya:2019dnh,Eichmann:2019bqf,Miramontes:2021exi} and the latest reviews~\cite{Aoyama:2020ynm,Colangelo:2022jxc} on the subject. 

\begin{figure*}[t!]
\centerline{%
\includegraphics[width=0.75\textwidth]{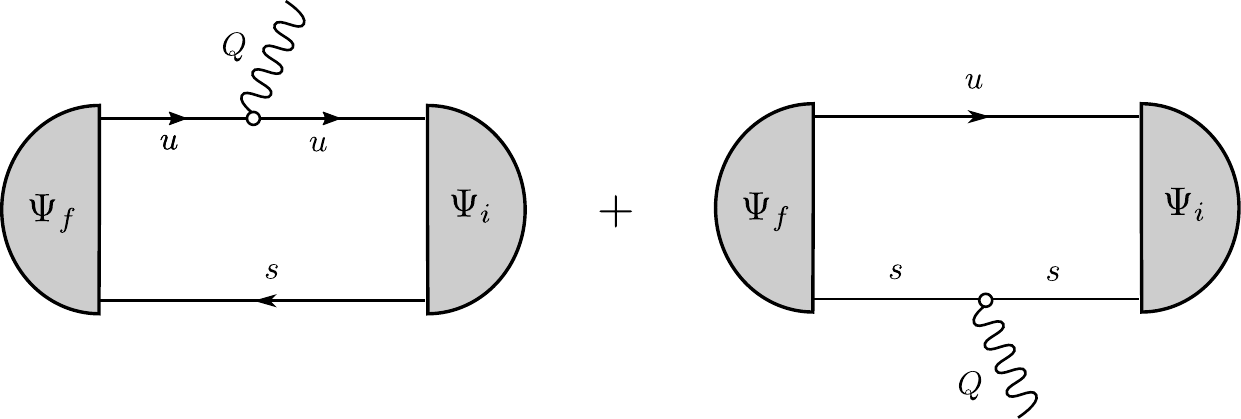}}
\caption{IA diagrams representing the coupling of the external field to the valence quarks. In the case of the electromagnetic charged kaon form factor, the external field couples to the $u$ (left diagram) and $s$ (right diagram) quarks. }
\label{fig:FF_IA}     
\end{figure*}

In this article, we exploit the same strategy as developed for a recent calculation of the time-like electromagnetic pion form factor~\cite{Miramontes:2021xgn} and study the electromagnetic $K^{\pm}$ form factor. In this case, the low energy time-like region is saturated not only by the $\rho$ resonance but also by the $\phi$ resonance around $m_{\phi} = 1.019$ GeV. Therefore, in order to capture the correct behaviour of the kaon form factor in this domain, the decay mechanism $\phi \rightarrow K K$ must incorporate additional features. To this end, we include explicit pions and kaon contributions into the Bethe-Salpeter interaction kernel and the quark propagators. 

The article is organized as follows, in 
Section~\ref{EFFs} we summarize the main elements of the SDE and BSE approach i.e, the quark propagators, the Bethe-Salpeter amplitudes (BSA), the quark-photon vertex (QPV) and the truncation scheme which goes {\em beyond} the usual {\em rainbow-ladder} (BRL) approximation employed for similar calculations. In section~\ref{NumericalR},  we present our numerical results for the kaon form factor in the time-like region using two different sets of parameters in the effective interaction. Finally, in section~\ref{Conclusions}, we present our conclusions and scope for future work.


\section{Electromagnetic form factors in the SDE/BSE formalism} \label{EFFs}

The interaction of a virtual photon with a pseudoscalar meson $M$ is described by a single form factor, $F_M(Q^2)$, and is conveniently written as:
\begin{equation}
\label{eq:FF}
\langle\textbf{P}(p_1)|J^\mu|\textbf{P}(p_2)\rangle= e(p_1 + p_2)^{\mu} F_M(Q^2)\;,
\end{equation}
 where $Q=p_1 -p_2$ is the probing photon four-momentum and $e$ is the elementary electromagnetic charge. On the other hand, the electromagnetic current, which describes the coupling of a single photon with a quark-antiquark system, can be expressed as
\begin{equation}
J^{\mu} = \bar{\Psi}_{\textbf{P}}^f G_0 (\mathbf{\Gamma}^{\mu} - K^{\mu}) G_0 {\Psi_\textbf{P}}^i ~,
\label{eq:Current}
\end{equation} 
where $\Psi_{\textbf{P}}^{i}$ and $\Psi_{\textbf{P}}^{i}$ are the BSA of the incoming and outgoing meson $M$, respectively, $G_0$ includes the appropriate product of dressed quark propagators and the term 
$\mathbf{\Gamma}^{\mu}$  represents the {\em impulse approximation} (IA) diagrams which represent the coupling of the photons to the dressed valence quarks (see Fig.~\ref{fig:FF_IA}),
\begin{equation}
\mathbf{\Gamma}^{\mu} = \left(S^{-1} \otimes S^{-1} \right)^{\mu} = \Gamma^{\mu}\otimes S^{-1} + S^{-1}\otimes \Gamma^{\mu}\; ,
\end{equation}
 where $S^{-1}$ is the inverse quark propagator. Effects beyond the IA terms are encoded in the second term involving $K^\mu$ which characterizes the interaction of the photon with the Bethe-Salpeter kernel describing the two-body interaction~\cite{Miramontes:2019mco,Miramontes:2021xgn}. Both the terms in Eq.~\eqref{eq:Current} need to be included in order to preserve charge conservation precisely. To compute the expression in Eq.~\eqref{eq:Current}, we require the knowledge of the quark propagators involved, the meson BSA, the QPV and the corresponding interaction kernels. To render this article
self-contained, we summarise  the key elements of the SDE/BSE approach in the next subsection, see recent and relevant reviews~\cite{Cloet:2013jya, Eichmann:2016yit, Huber:2018ned, Sanchis-Alepuz:2017jjd, Bashir:2012fs} for further details.

\begin{figure*}[t!]
\centerline{%
\includegraphics[width=0.95\textwidth]{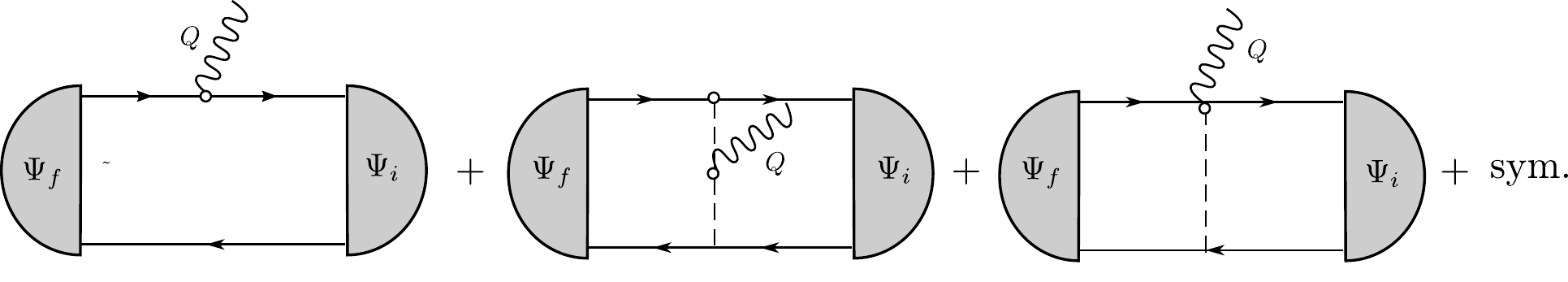}}
\caption{Beyond IA diagrams necessary for a consistent calculation when we include meson exchange into the interaction kernel. The left diagram corresponds to the IA diagram, the middle diagram represents the coupling of the photon to the meson and the last diagram the coupling of the photon with a meson-quark vertex. For this work, we have included only the first IA diagram.}
\label{fig:current_BIA}     
\end{figure*}

\subsection{Elements of the SDE/BSE approach}
The coupled SDE/BSE formalism to calculate form factors needs basic ingredients which we discuss below. Note that all computation which follows is carried out in the Euclidean momentum space.\\

\noindent {\bf{The quark-photon vertex:}}
Due to the fortunate coincidence that the strongly interacting quarks are also electrically charged, the internal structure of a meson can be probed through a photon interacting with its constituents. The fully-dressed QPV, $\Gamma^{\mu}$, which characterizes the interaction between quarks and photons in quantum electrodynamics, can be described via the following inhomogeneous BSE:
\begin{eqnarray}
\label{eq:inhomBSE_vector}
\bracket{\Gamma^{\mu}}_{a\alpha,b\beta}\bracket{p,Q}&=&
Z_2 \bracket{\gamma^\mu}_{ab} t_{\alpha\beta}\\
&+&\int_q K^{r\rho,s\sigma}_{a\alpha,b\beta}\bracket{Q,p,q} \, S_{r\rho,e\epsilon}\bracket{k_1}
\nonumber \\
&\times& \bracket{\Gamma^{i,\mu}}_{e\epsilon,n\nu}\bracket{Q,q}S_{n\nu,s\sigma}\bracket{k_2} \,,
\nonumber
\end{eqnarray}
where $Z_2$ is the quark wave-function renormalization constant and $S(k)$ is the fully dressed quark propagator. Among the kinematic factors, $Q$ is the photon momentum, $p$ and $q$ are the external and internal relative momenta between the quark and the antiquark, respectively, while the internal quark and antiquark momenta are defined as $k_1=q+Q/2$ and $k_2=q-Q/2$ such that $Q=k_1-k_2$ and $q=(k_1+k_2)/2$. The Latin  letters represent Dirac indices while the Greek letters represent flavour indices. The isospin structure of the vertex is given by 
$t_{\alpha\beta}= \textrm{diag}\bracket{\nicefrac{2}{3},\nicefrac{-1}{3}, \nicefrac{-1}{3}}$. The QPV can be decomposed in a basis composed of eight vectors transverse to the photon momentum and four non-transverse ones, all of which are included in our calculation. \\

\noindent {\bf{The Bethe-Salpeter Equation:}}
The description of mesons as bound states of two quarks in our framework is effectuated through the homogeneous BSE
\begin{eqnarray}
\label{eq:homogeneousBSE}
\bracket{\Gamma}_{a\alpha,b\beta}\bracket{p,P}= \int_q K^{r\rho,s\sigma}_{a\alpha,b\beta}\bracket{P,p,q} \times S_{r\rho,e\epsilon}\bracket{k_1} \nonumber\\
\bracket{\Gamma}_{e\epsilon,n\nu}\bracket{q,P}S_{n\nu,s\sigma}\bracket{k_2}~,
\end{eqnarray}
\noindent where $P$ is the total momentum of the meson. For pseudoscalar mesons, abbreviated as $_{PS}$ in the subscripts below, the Dirac part of the BSA $\Gamma$ can be expanded in a tensorial basis with four elements:
    \begin{eqnarray}
&& \hspace{-4mm} \Gamma^i_{PS}(p,P) =\tau^i~\gamma_5 \{ E_{PS}(p,P) - i \slashed{P} F_{PS}(p,P) \nonumber \\
 && \hspace{12mm} - i\slashed{p} (p \cdot P) G_{PS}(p,P) - \left[ \slashed{P},\slashed{p}\right]H_{PS}(p,P) \} \,,~\label{eq:pionamplitude}
\end{eqnarray}
where $E_{PS},F_{PS} ,G_{PS}$ and $H_{PS}$ are four independent dressing functions. The two-body interaction in Eq.~\eqref{eq:homogeneousBSE} is represented by $[K(P,p,q)]$; it corresponds to two-particle irreducible quark/antiquark scattering kernel, which  contains all possible interactions between the quark and the antiquark within the bound state. \\

\noindent {\bf{The quark propagator:}}
The dressed quark propagator can be obtained  by solving the corresponding SDE, which reads 
\begin{equation}
S^{-1} = S_0^{-1} - Z_{1f} \int_q \gamma_\mu S(q) \Gamma^{qgl}_\nu(q,k) D_{\mu \nu}(k)~,
\label{eq:quarkDSE}
\end{equation}
where $S_0^{-1}(p)$ the renormalized bare propagator,
\begin{equation}
S_0^{-1}(p) = Z_2\left(i \slashed{p} + Z_m m\right)~,
\end{equation}
and $Z_{1f}$, $Z_2$ and $Z_m$ are renormalization constants associated with the quark-gluon vertex, the quark propagator and the quark mass, respectively. Moreover, $m$ is the  current quark mass, $\Gamma^{qgl}$ is the full quark-gluon vertex and $D_{\mu \nu}$ is the full gluon propagator which, in the Landau gauge acquires the following form
\begin{equation}
D_{\mu \nu}(k) = \left( \delta_{\mu \nu} - \frac{k_\mu k_\nu}{k^2}\right) \frac{Z(k^2)}{k^2}~,
\end{equation}
with $Z(p^2)$ being the gluon dressing function. For the simplicity of notation, we have suppressed the color indices.

\subsection{Rainbow ladder truncation and beyond}
\label{sec:RL}
In the SDE/BSE formalism, one of the simplest non-perturbative approximation that preserves  the axial-vector (AxWTI) and vector Ward-Takahashi identities (VWTI) is the so called rainbow-ladder (RL) truncation, where the BSE interaction kernel is simplified to a vector-vector gluon exchange with an effective coupling $\alpha\bracket{k^2}$. Omitting colour indices, this kernel can be written as  
\begin{equation}
K^{r\rho,s\sigma}_{a\alpha,b\beta}\bracket{Q,p,q}=\alpha\bracket{k^2}\gamma^\mu_{ar}\gamma^\nu_{sb}D^{\mu\nu}\bracket{k}\delta^{\alpha\rho}\delta^{\sigma\beta}~,\label{eq:RLkernel}
\end{equation}
where $k=p-q$ is the momentum flowing through the gluon propagator. On the other hand, the truncated quark SDE in the RL truncation reads,
\begin{equation}
Z_{1f} \gamma_{\mu} Z(k^2)\Gamma_{\nu}^{\text{qgl}}(q,p)  \rightarrow Z_2^2 \gamma_{\mu} 4\pi \alpha(k^2) \gamma_{\nu} \,.
\end{equation}
Note that the effective coupling $\alpha(k^2)$ describes the strength of the quark-antiquark interaction.
There exist primarily two models  in the literature to parameterize it: the Maris-Tandy (MT) model~\cite{Maris:1997tm,Maris:1999nt} and the Qin-Chang (QC) model \cite{Qin:2011dd}.\footnote{It is worth mentioning that both the models (MT and QC) for the effective coupling $\alpha(k^2)$ yield very similar results for hadron spectroscopy and  structure. We choose to employ the MT model in order to make consistent comparison with the work reported earlier on the time-like pion electromagnetic form factor~\cite{Miramontes:2021xgn}.}  For  the purpose of this article, we employ the MT model. It reads as follows
\begin{eqnarray}
\alpha(q^2) &=&
 \pi\eta^7\left(\frac{q^2}{\Lambda^2}\right)^{2}
e^{-\eta^2\frac{q^2}{\Lambda^2}} \nonumber \\
&+&\frac{2\pi\gamma_m
\big(1-e^{-q^2/\Lambda_{t}^2}\big)}{\textnormal{ln}[e^2-1+(1+q^2/\Lambda_{QCD}
^2)^2]} \,, \label{eq:MTmodel}
\end{eqnarray}
which includes a Gaussian term, dominant in the infrared, to provide enough interaction strength for DCSB to take place and an ultraviolet dominant term which reproduces the one-loop QCD behavior of the quark propagator at large momenta. The free parameters of the MT model are  $\Lambda$ and $\eta$. These are usually fitted to reproduce the pion mass and the its weak decay constant. In the SDE/BSE framework, the running quark masses $m_u$, $m_d$ and $m_s$ are also introduced as input parameters. The scale $\Lambda_t=1$~GeV is brought in for technical reasons alone and has no bearing on the computed observables.  The anomalous dimension is $\gamma_m=12/(11N_c-2N_f)=12/25$ with $N_f=4$ flavours and $N_c=3$ colours. For the QCD mass scale, we use $\Lambda_{QCD}=0.234$ GeV.

In the RL truncation, the solutions of the homogeneous BSE are bound sates which cannot develop a decay width. Note that the bound states, determined as solutions of BSEs, appear as poles in Green functions with the corresponding quantum numbers. In particular, when we compute the QPV in the RL truncation, electrically neutral vector mesons manifest themselves as poles located on the negative real axis of $Q^2$. As a consequence, the electromagnetic form factor develops the same pole on the real axis of the time-like region. However, in general, any dynamics that is correlated with the presence of virtual intermediate particles, {\it e.g.}, decays, will be absent from the results of any calculation using only the RL truncation. Therefore, in order to shift the poles from the negative real axis to the complex plane to incorporate decay mechanisms, we employ the BRL truncation. It is enforced by incorporating the presence of intermediate particles in the interaction kernel, leading to an adequate description of the time-like form factors. 

In Refs.~\cite{Fischer:2007ze, Fischer:2008sp, Sanchis-Alepuz:2014wea}, the authors implemented the appearance of intermediate particles by introducing explicit pionic degrees of freedom and pion-quark interactions both into the truncated SDE and in the interaction kernel. The pion-quark interaction vertex, represented by the pion BSA is calculated via a consistently truncated BSE. The explicit and illustrative diagrammatic modification of the quark SDE as well as the interaction kernel are presented in Appendix~\ref{kernels}. In addition to the gluon exchange, this BRL truncation contains  a $t-$channel meson exchange (second term in lowest row of Fig.~\ref{fig:kernels}) and an $s-,u-$channel meson exchange (third and fourth terms in the lowest row of Fig.~\ref{fig:kernels}). At this point, it is informative to recall that the authors in~\cite{Fischer:2007ze,Fischer:2008sp,Sanchis-Alepuz:2014wea} only take into account the gluon exchange plus the $t-$channel meson exchange diagrams.  
The implementation of the $s-,u-$channel meson exchange into the interaction kernel leads to several numerical difficulties, such as contour deformation integration to calculate the BSE, which are detailed in Refs.~\cite{Miramontes:2019mco, Williams:2018adr}. Our BRL truncation yields a correct description of the QPV which develops a multi-particle branch cut along the negative real $Q^2$-axis, starting at the two-pion production threshold as expected~\cite{Miramontes:2019mco}. Additionally, it allows for the description of the $\rho$-meson as a finite-width resonance which is primarily due to the intermediate process $\rho \rightarrow \pi \pi$~\cite{Williams:2018adr}. Subsequently, the electromagnetic pion form factor was calculated in~\cite{Miramontes:2021xgn}. The absolute value and the phase of this physical observable reported there compare favorably to the available experimental data for the space-like and the time-like momenta, developing the expected ``bump" on the negative real $Q^2$-axis. In~\cite{Miramontes:2021exi}, we calculated pion and kaon space-like electromagnetic form factors and computed box contributions to the hadronic light-by-light piece of the muon’s anomalous magnetic moment $a_{\mu}$, where our results are compatible with the ones reported using other reliable contemporary frameworks. Finally, in~\cite{Miramontes:2022mex}, this truncation was employed to study isospin symmetry breaking and its effect on the light meson masses.

For the study of the electromagnetic charged kaon form factor, we follow the same strategy as the one adopted in the previous  work on the electromagnetic pion form factor \cite{Miramontes:2021xgn}. The mass difference between  $u$ and $s$ quarks, which compose the charged kaon, leads to an interference between different vector meson resonances. In this case, in addition to the $\rho$ meson resonance, a description of the $\phi$ meson and its decay $\phi \rightarrow K K$ needs to be properly implemented into the interaction kernel. To include the decay mechanism, kaon emission and absorption is included, together with a kaon-quark interaction, given by the full kaon BSA, into the strange quark SDE. The interaction kernel now includes the exchange of kaons in the $t-,s-$ and $u-$channels (see Fig.~\ref{fig:kernels}). The IA diagrams necessary to calculate the electromagnetic current are depicted in Fig.~\ref{fig:current_BIA}. While the first diagram (the coupling of the photon with the $u$ quark) encodes the decay $\rho \rightarrow \pi \pi$ via pionic correction in the up-quark SDE and pion exchange in different channels in the interaction kernel, the second diagram (the coupling of the photon to the $s$ quark) includes the decay $\phi \rightarrow K K$ via kaonic corrections in the strange-quark SDE and kaon exchange in different channels in the interaction kernel. 

We wish to reemphasize here, as already discussed in~\cite{Miramontes:2021xgn}, this BRL truncation does not satisfy the AxWTI and VWTI simultaneously. We can choose to satisfy one of the two identities while mildly violating the other one. The AxWTI identity ensures that pion is a massless bound state in the chiral limit, while the VWTI leads to charge conservation in electromagnetic processes. Since we  focus on electromagnetic form factors, we choose to preserve the VWTI and ensure charge conservation. Nevertheless, we observe that the violation of the AxWTI leads only to a small and insignificant error in relevant physical observables.

It is also worth mentioning that in the previous calculation of the electromagnetic pion form factor, the meson vertices appearing in the interaction kernel, Eqs.~(\ref{eq:BSEkernel_tchannel}-\ref{eq:BSEkernel_uchannel}), were approximated by the leading pion BSA proportional to $\gamma_5$, whereas in this current calculation for the kaon electromagnetic form factor, the full BSA has been considered, leading to a better description of the $\rho$ resonance and its decay width, as detailed in the following section.  

\section{Numerical Results} \label{NumericalR}

\begin{figure*}[t!]
\centerline{%
\includegraphics[width=1.0\textwidth]{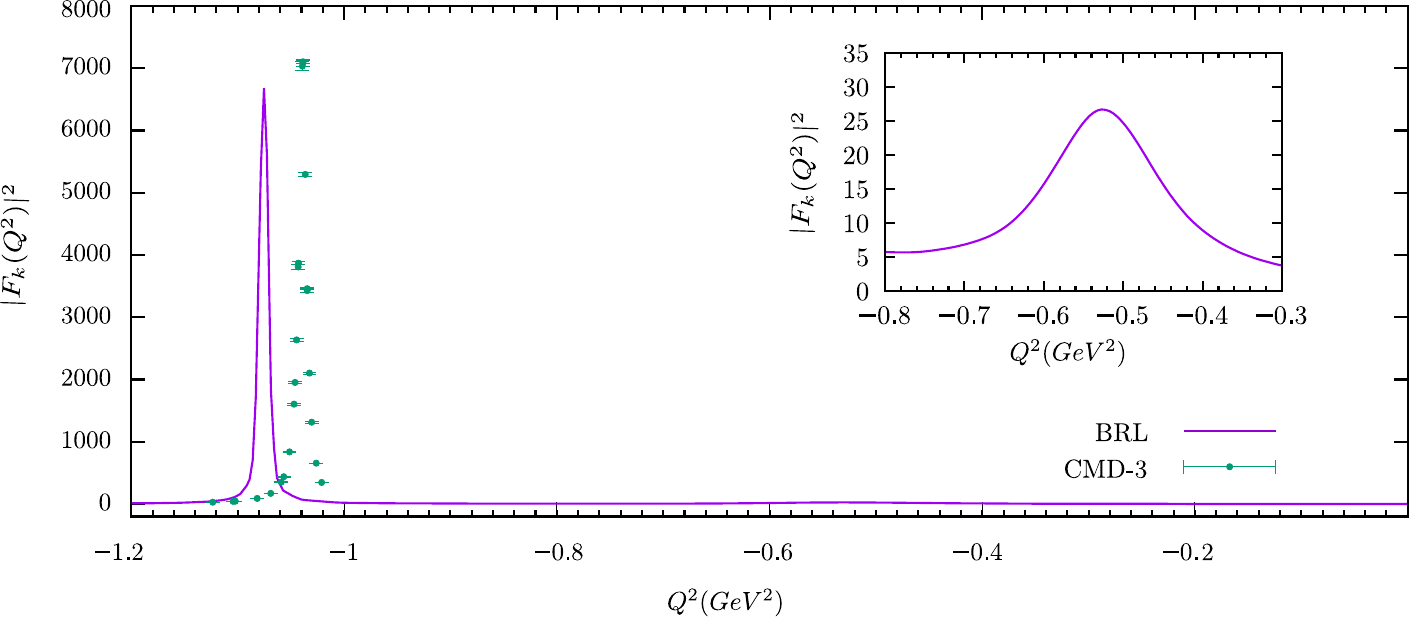}}
\caption{Absolute value squared of the electromagnetic kaon form factor on the time-like $Q^2 <0 $ region. We use the parameters $\eta = 1.6$ and $\Lambda = 0.74$ GeV. We provide a comparison with experimental measurements extracted from~\cite{Kozyrev:2017agm}. Both resonances ($\rho$ and $\phi$) are reproduced in our calculation.}
\label{fig:FF_kaon_full}     
\end{figure*}
Following the SDE/BSE framework described in the previous section, we calculate the electromagnetic charged kaon form factor in the time-like range $Q^2 < 0 $. We work in the $SU(2)$ isospin symmetric limit, where mass of the up and down quarks are taken to be equal. We adjust the free parameters $\eta$ and $\Lambda$ in the effective MT interaction, together with the quark masses $m_u$, $m_d$ and $m_s$, in order to reproduce the experimental values of the mass and decay constant for the pion and kaon mesons. We employ two different sets of parameters in the MT interaction: 
\begin{itemize}
 \item BRL-I: $\eta = 1.5$  $\Lambda = 0.78$ GeV.
 \item BRL-II: $\eta = 1.6$,  $\Lambda = 0.74$ GeV.
 \item For the current quark masses, we employ the values $m_{u/d} = 6.8$ MeV and $m_s = 85.0$ MeV at the renormalization scale $\mu = 19$ GeV.
\end{itemize}
  It is important to bear in mind that in order to have a fully consistent computational framework, the electromagnetic current has to be calculated in the beyond-IA scheme employed, including diagrams of the photon coupling with the corresponding meson and with a meson-quark vertex (see Fig.~\ref{fig:current_BIA}). Nevertheless, to alleviate cumbersome numerical effort, we compute only the IA diagrams.

In Table I, we  collect the fitted masses for the pion and the kaon as well as the predicted masses for the $\rho$ and the $\phi$ mesons, taking into account only the RL truncation plus the $t$-channel meson exchange in the interaction kernel (i.e. without the $s-$ and the $u-$-exchange channels). Additionally, we include the results for different masses calculated by employing only the RL truncation with the values $\eta = 1.8$ and $\Lambda = 0.72$ for comparison. Note that for each case, we also report the weak decay constants. From Table I, we can readily extract informative conclusions:
\begin{itemize}

\item Employing only the RL truncation with the parameters that reproduce the correct values of the pion and kaon masses, the $\phi$ meson is around 50 MeV above the experimental value, while the $\rho$ meson is around 45 MeV below the experimental value. It seems to suggest that it is impossible to obtain the correct $\rho$ and $\phi$ masses simultaneously with one set of parameters in the MT interaction in our approach. 

\item Nevertheless, in this article we aim to describe primarily the qualitative behaviour of the electromagnetic form factor in the time-like region. Therefore, we will consider a 5$\%$ deviation from the experimental data sufficiently acceptable in one of the first efforts made in this direction through the SDE/BSE formalism.

\item

The set BRL-I reproduces a better value for the $\rho$ mass, but at the cost of a larger value for the $\phi$ meson mass. The set BRL-II is adopted in order to get a more acceptable result for the $\phi$ mass but this is achieved at the expense of an undesirable reduction in the masses of the $\pi$, $K$ and $\rho$ mesons. The upshot of this exercise is that we seek the best compromise with the truncation schemes hitherto available to us. 

\end{itemize}


\begin{table}[]
 \centering
\begin{tabular}{|S|SSSS|}
\hline
  {}         & {RL}    &   {BRL I} & {BRL II} & {Exp.}  \\ \hline
{$m_{\pi}$}      & {0.140} & {0.139} & {0.128}  & {0.139} \\
{$f_{\pi}$}      & {0.131} & {0.138} & {0.135}  & {0.131} \\ 
{$m_{_k}$}     & {0.493} & {0.493} & {0.480}  & {0.493} \\ 
{$f_k$}       & {0.155} & {0.159} & {0.157}  & {0.156} \\ 
{$m_{\rho}$}     & {0.730} & {0.762} & {0.737}  & {0.775} \\ 
{$f_{\rho}$}     & {0.215} & {0.225} & {0.220}  & {0.216} \\ 
{$m_{\phi}$}     & {1.070} & {1.077} & {1.036}  & {1.019} \\ 
{$f_{\phi}$}     & {0.259} & {0.265} & {0.260}  & {0.236} \\\midrule
{$M_{\rho}$}     & {0.739} & {0.752} & {0.728}  & {0.775} \\ 
{$\Gamma_{\rho}$} & {0.0}  & {0.122} & {0.123}  & {0.145} \\ 
{$M_{\phi}$}     & {1.070} & {1.077} & {1.030}  & {1.019} \\ 
{$\Gamma_{\phi}$} & {0.0}   & {0.006} & {0.005}  & {0.004} \\\bottomrule
\end{tabular}
\caption{Numerical results for the masses ad decay constants. In the RL we use $\eta=1.8$ and $\Lambda = 0.72$ GeV, for BRL 1 we employ $\eta=1.5$ and $\Lambda = 0.78$ GeV, and for BRL II $\eta=1.6$ and $\Lambda = 0.74$ GeV. The masses $M_\rho$, $M_{\phi}$ and the respective decay widths have been extracted from our resulting time-like form factor using Pade approximant. All mass-dimensioned quantities are expressed in GeV. The experimental values have been extracted from Ref.~\cite{10.1093/ptep/ptac097}.}
\end{table}

\begin{figure*}[t]
\centerline{%
\includegraphics[width=0.76\textwidth]{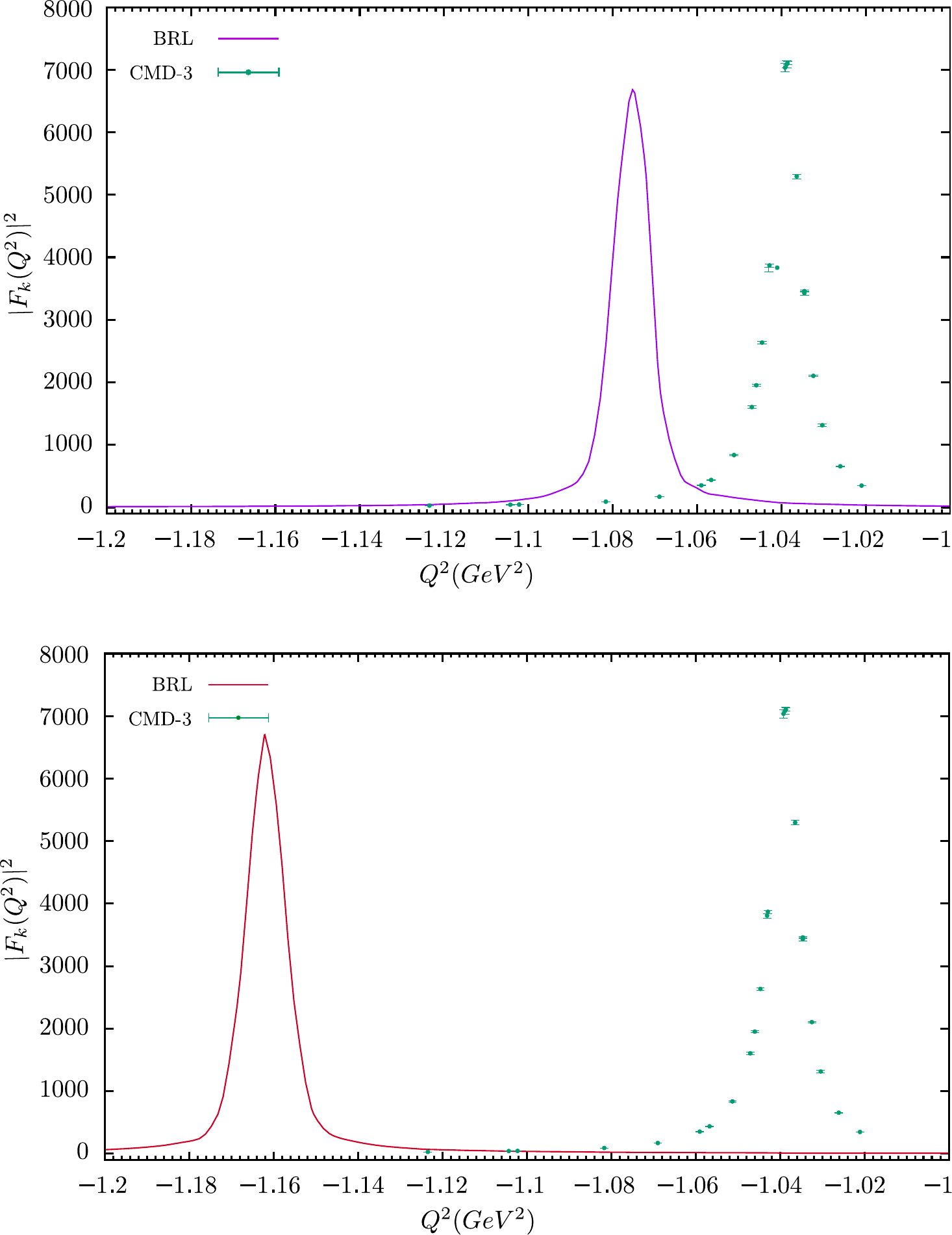}}
\caption{Absolute value squared of the time like kaon form factor using different values for the free parameters in the MT interaction, the upper panel: $\eta = 1.5$, $\Lambda = 0.78$ and the bottom panel: $\eta = 1.6$, $\Lambda = 0.74$ }
\label{fig:FF_kaon}     
\end{figure*}

In Figs.~\ref{fig:FF_kaon_full} and \ref{fig:FF_kaon} we show our results for the kaon electromagnetic form factor for a time-like ($Q^2<0$) photon. As already discussed in the text, there are two main approximations within the scope of our calculation. 

\begin{itemize}
    \item 
    The first is the IA, only taking into account the diagrams where the photon couples to the $u$ and $s$ quarks (Fig.~\ref{fig:FF_IA}), neglecting further couplings. It leads to a deviation from $F_k(Q^2 = 0) = 1$ of around only 1$\%$.
    \item
    As already mentioned earlier in the text, the second approximation relates to including intermediate resonances into the interaction kernel. It entails that one must choose whether the AxWTI or the VWTI is preserved, while the other is violated. Since we are interested in the electromagnetic form factor, the symmetry we naturally want is the conservation of charge. Therefore, we choose to ensure the VWTI is satisfied throughout.
\end{itemize}

As already demonstrated in \cite{Miramontes:2019mco}, the inclusion of different meson exchange channels into the interaction kernel leads to a branch cut which lies along the real axis for $Q^2<0$, starting at the two-particles production threshold $Q^2=-4m_{\pi,K}^2$ \footnote{As detailed in \cite{Miramontes:2019mco}, in addition to the multiparticle branch cuts, the quark propagators also manifest themselves as branch cuts that we have to deal with while integrating to solve for the QPV. For the two different sets of MT parameters we are able to calculate the form factor up to around -1.5 GeV$^2$. Beyond this value, the branch cuts from the pions/kaons and the branch cuts from the quark propagators begin to overlap and the contour deformation is no longer possible for the required integration. \\}. The time-like electromagnetic form factors thus acquire the same branch cuts. Therefore, they are defined as $F(Q^2 + i \epsilon)$. For the present calculation we use $\epsilon = 0.0001$ GeV$^2$. The inclusion of the decay kernels moves the $\rho$ and $\phi$ resonances poles from the real axis to the complex plane in the QPV, as is manifest in Figs.~\ref{fig:FF_kaon_full} and~\ref{fig:FF_kaon}. In Fig.~\ref{fig:FF_kaon_full}, we depict our results for the kaon electromagnetic form factor with the BRL-1 truncation from 0.0 GeV$^2$ to -1.2 GeV$^2$. The extracted $\phi$ resonance pole in this case lies at the mass value $M = 1.030$ GeV and has the decay width of $\Gamma_\phi = 0.005$ GeV. 
Additionally, the corresponding kaon charge radii obtained are $\sqrt{r^2_K} = 0.592$ fm for BRL-I and $\sqrt{r^2_K} = 0.590$ for BRL-II.

In the inset of Fig.~\ref{fig:FF_kaon_full} we can observe that the $\rho$ pole is also present even though it is not apparently visible on the main plot~\footnote{It is important to note that the radially excited states such as $\rho(1400)$ are also encoded in our truncation. Nevertheless, the corresponding resonance poles appear outside the range of our calculation. Results on the radially excited states employing the SDE/BSE formalism will be published elsewhere.}. In Fig.~\ref{fig:FF_kaon} we show the electromagnetic kaon form factor in the region $[-1, -1.2]$ GeV$^2$, where we can notice that the resonance pole position is sensitive to the parameters which define the MT interaction. We wish to emphasize again that including only RL plus the t-channel mesons exchange into the interaction kernel implies that the $\rho$ and $\phi$ mesons are stable bound states and do not develop a decay width, i,e. the poles are located in the negative real axis of the time-like form factor. It is the inclusion of both the $s-$ and $u-$channels meson exchange,  with not widely disparate contributions, which shifts the resonance poles from the real axis to the complex plane, describing the correct behaviour of the form factor. We parameterize the solution of the electromagnetic kaon form factor employing a Padé fit in order to extract the pole position as $M^2_{{\rm pole}} = - Q^2 + i M_{{\rm pole}} \Gamma$, where $M_{{\rm pole}}$ is the mass of the meson and $\Gamma$ is the corresponding decay width. In Table~I, we collect the values of the extracted masses and decay widths for the $\rho$ and $\phi$ mesons. In our previous calculation of the electromagnetic pion form factor in Ref.~\cite{Miramontes:2021xgn} we observed that using only the leading component $\gamma_5$ of the pion BSA into the vertices of the interaction kernel, underestimates the decay width of the $\rho$ to $\Gamma = 0.105$ GeV. Using the full BSA in the vertices, leads to a much better result of $\Gamma = 0.123$ GeV. 

 Additionally we explicitly compute the separate contributions arising from the $s-$ and $u-$ meson exchange channels on the total quantitative outcome. We conclude that the $s-$ channel is relatively more dominant. It contributes around 65\% to the decay widths, while the remaining contribution stems from the $u-$ meson exchange channel. Therefore none can be ignored in terms of their quantitative impact. Moreover, it is worth emphasizing that both the channels must be included in order to satisfy the Ward identities.

\section{Conclusions and scope} \label{Conclusions}

In this article, we present an exploratory study of the electromagnetic charged kaon form factor in the time-like region employing the formalism of SDE/BSE. We focus on the effects of intermediate mesons (pions and kaons) in the interaction kernel. The main approximations we employ in this calculation are the IA and the small and numerically insignificant violation of the AxWTI. Notwithstanding, the results obtained are in fairly satisfactory agreement with the experimental observations. Our results are consistent with the fact that the kaon form factor is not only saturated by the $\rho(770)$ resonance, but mainly by the $\phi(1020)$ resonance in the appropriate low energy regime. As can be deciphered in the plots illustrated in the previous section, our results develop two \textit{bumps} close to the experimental mass values for the $\rho$ and $\phi$ mesons. The exact pole position is sensitive to the choice of the free parameters in the MT interaction. The decay width extracted from our results for the case of the $\rho$ shows a clear improvement on our previous calculation. In the present case, we have employed the full pion BSA into the pion vertices appearing in the system of SDE/BSE. On the other hand, our result for the decay width for the $\phi$ meson is very close to experimental results. It is worth mentioning that our calculation has been carried out in the isospin symmetric limit. Thus it lacks $\omega-\rho$ mixing effects. These effects might play a noticeable role in capturing the correct behaviour of the time-like form factor, especially close to the resonances. First results on isospin breaking effects on the splitting of light mesons have been explored using the SDE/BSE framework in~\cite{Miramontes:2022mex}. In the future, we would like to implement this effects into our calculations of form factors.  \\

\begin{acknowledgments}
A.~S.~Miramontes acknowledges {\em Consejo Nacional de Ciencia y Teconolog\'ia} (CONACyT), Mexico, for the financial support provided to him through the program ``{\em Postdoctorados Nacionales por M\'exico}". A.~Bashir is grateful to the {\em Universidad Michoacana de San Nicol\'as de Hidalgo}, Morelia, Mexico, for the {\em Coordinaci\'on de la Investigaci\'on Cient\'ifica} (CIC) project grant no. 4.10. 

\end{acknowledgments}

\vfil\eject


\appendix
\section{Interaction kernels}
\label{kernels}

\begin{figure*}[ht]
\centerline{%
\includegraphics[width=0.95\textwidth]{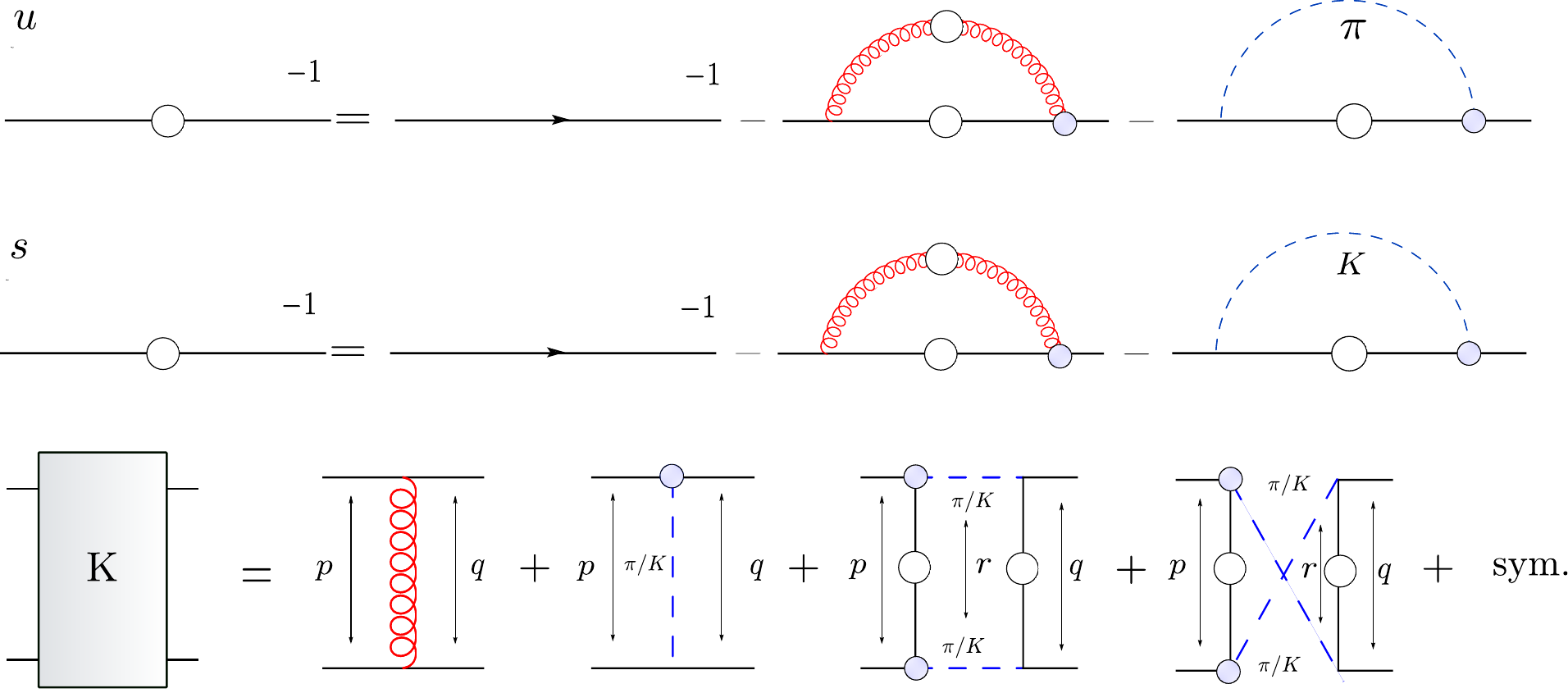}}
\caption{Truncations used herein for the BSE interaction kernel $K$ (lower diagram) and the quark SDE one (upper diagram). 
In the lower diagram, the terms on the right-hand side correspond to the rainbow-ladder, pion/kaon exchange, and the $s$- and the $u$-channel pion/kaon decay contributions to the truncation, respectively. The $s$- and the $u$-channel pion/kaon decay terms do not contribute to the 
quark SDE.}
\label{fig:kernels}     
\end{figure*}
The Bethe-Salpeter interaction kernel which represents the exchange of explicit mesonic degrees of freedom as defined in
\cite{Fischer:2007ze, Fischer:2008sp} reads as follows,
\begin{flalign}
&K^{(t)~ut}_{rs}(q,p;P) =\nonumber\\
&~~~~~~~\frac{C}{4} [\Gamma_{\textbf{P}}^j]_{ru} \left(\frac{p + q - P}{2}; p - q \right) [Z_2 \gamma^5]_{ts} D_{\textbf{P}}(p - q) \nonumber \\ \nonumber
 &~~~+\frac{C}{4} [\Gamma_{\textbf{P}}^j]_{ru} \left(\frac{p + q - P}{2}; q - p \right) [Z_2 \gamma^5]_{ts} D_{\textbf{P}}(p - q) \\ \nonumber
 &~~~+\frac{C}{4} [Z_2 \gamma^5]_{ru} [\Gamma_{\textbf{P}}^j]_{ts} \left(\frac{p + q + P}{2}; p - q \right) D_{\textbf{P}}(p - q) \\ 
 &~~~+\frac{C}{4} [Z_2 \gamma^5]_{ru} [\Gamma_{\textbf{P}}^j]_{ts} \left(\frac{p + q + P}{2}; q - p \right) D_{\textbf{P}}(p - q)~,\label{eq:BSEkernel_tchannel}
\end{flalign}
\noindent in combination with the following truncation of the quark SDE
\begin{eqnarray}
S^{-1}(p) &=& S^{-1}(p)^{RL} - \frac{3}{2} \int_q \Bigg[Z_2 \gamma_5 S(q) \Gamma_{\pi}\left(\frac{p+q}{2}, q-p\right) \nonumber \\
&&  + Z_2 \gamma_5S(q)\Gamma_{\textbf{P}}\left(\frac{p+q}{2}, p-q\right)\Bigg] \frac{D_{\pi}(k)}{2}~,\label{eq:quarkDSE_tchannel}
\end{eqnarray}
with \textbf{P} the corresponding $\pi$, $K$ meson and $S^{-1}(p)^{RL}$ being the right-hand-side of the quark SDE in the RL truncation with the gluon-mediated interaction as described in Sect.~\ref{sec:RL}. 
In Eqs.~\eqref{eq:BSEkernel_tchannel} and \eqref{eq:quarkDSE_tchannel} the meson propagator is taken as 
$D_{\pi}(k) = {(k^2 + m_{\pi}^2)^{-1}}$. 

Herein, the quark-meson vertex $\Gamma_\pi$ is taken to be the full pion/kaon BSA. The exchange of the mesons in the kernel can also appear in the $s$- and the $u$-channels~\cite{Fischer:2007ze}. They read:


\begin{widetext}
\flaligne{
K^{(s)~he}_{da}(q,p,r;P)=~& \frac{C}{2}~D_{\textbf{P}}\left(\frac{P + r}{2}\right) D_{\textbf{P}}\left(\frac{P - r}{2}\right)~\left[[Z_2\gamma_5]_{dc} S_{cb}\left(p - \frac{r}{2}\right)[Z_2\gamma_5]_{ba}\right.\nonumber\\
&~~~~\times[\Gamma_{\textbf{P}}^j]_{hg} \left(q - \frac{P}{4} - \frac{r}{4}; \frac{r - P}{2} \right) S_{gf}\left(q - \frac{r}{2}\right)[\Gamma_{\textbf{P}}^j]_{fe} \left(q + \frac{P}{4} - \frac{r}{4}; -\frac{P + r}{2} \right)\nonumber\\
~&+~[\Gamma_{\textbf{P}}^j]_{dc} \left(p + \frac{P}{4} - \frac{r}{4}; \frac{P + r}{2} \right) S_{cb}\left(p - \frac{r}{2}\right)[\Gamma_{\textbf{P}}^j]_{ba} \left(p - \frac{P}{4} - \frac{r}{4}; \frac{P - r}{2} \right)\nonumber\\
&~~~~\left.\times[Z_2\gamma_5]_{hg} S_{gf}\left(q - \frac{r}{2}\right)[Z_2\gamma_5]_{fe}\right]~, 
\label{eq:BSEkernel_schannel}}

and

\flaligne{
K^{(u)~he}_{da}(q,p,r;P)=~& \frac{C}{2}~D_{{\textbf{P}}}\left(\frac{P + r}{2}\right) D_{{\textbf{P}}}\left(\frac{P - r}{2}\right)~\left[ [Z_2\gamma_5]_{dc}  S_{cb}\left(p + \frac{r}{2}\right) [Z_2\gamma_5]_{ba}\right.   \nonumber \\
 &~~~~\times [\Gamma_{\textbf{P}}^j]_{hg} \left(q - \frac{P}{4} - \frac{r}{4}; \frac{r - P}{2} \right) S_{gf}\left(q - \frac{r}{2}\right) [\Gamma_{\textbf{P}}^j]_{fe} \left(q + \frac{P}{4} - \frac{r}{4}; -\frac{P + r}{2} \right)\nonumber\\
~&+~[\Gamma_{\textbf{P}}^j]_{dc} \left(p + \frac{P}{4} + \frac{r}{4}; \frac{P - r}{2} \right) S_{cb}\left(p + \frac{r}{2}\right)[\Gamma_{\textbf{P}}^j]_{ba} \left(p - \frac{P}{4} + \frac{r}{4}; \frac{P + r}{2} \right)  \nonumber \\
 &~~~~\left.\times[Z_2\gamma_5]_{hg} S_{gf}\left(q - \frac{r}{2}\right)[Z_2\gamma_5]_{fe}\right]~,
\label{eq:BSEkernel_uchannel}
}
\end{widetext}
where $r$ is an additional integration momentum in the BSE (cf. Eqs.~\eqref{eq:inhomBSE_vector} or~\eqref{eq:homogeneousBSE}). The resulting truncation of the BSE kernel as well as the quark SDE are depicted in Fig.~\ref{fig:kernels}.
The inclusion of the $s$ and $u$-channel meson exchange into the interaction kernel given in equations (\ref{eq:BSEkernel_schannel}) and (\ref{eq:BSEkernel_uchannel}) generates a highly non-trivial analytic structure of the integrals in the BSE. It is induced by the intermediate pions/kaons potentially going on-shell as well as by singularities in the quark propagators, see Ref.~\cite{Miramontes:2019mco} for details. The techniques for finding viable contour deformations for performing the numerical integration in a mathematically correct way are also detailed therein.

\section{Poles of the quark propagator}\label{sec:quark_poles}

It is well known that the truncated quark SDE employed in this article features pairs of complex conjugate poles in the complex plane  (see, e.g., Ref.~\cite{Windisch:2016iud}). In order to facilitate the use of the quark propagators and easily identify the analytic structures generated by those poles in the form factor as well as vertex calculations, it is useful to parameterize the quark propagator simply as a sum of two terms containing complex conjugate poles (see, e.g., Ref.~\cite{El-Bennich:2016qmb}) 
\begin{eqnarray}
S(p) &=& -i \slashed{p} \sigma_v(p^2) + \sigma_s(p^2) \, ,\nonumber \\
\sigma_v (p^2) &=& \sum_{i}^n \left[\frac{\alpha_i}{p^2 + m_i} + \frac{\alpha_i^\ast}{p^2 + m_i^\ast}\right]  \, ,\nonumber \\ 
\sigma_s (p^2) &=& \sum_{i}^n \left[\frac{\beta_i}{p^2 + m_i} + \frac{\beta_i^\ast}{p^2 + m_i^\ast}\right]~,
\end{eqnarray}
where the parameters $m_i$, $\alpha_i$, $\beta_i$ can be obtained by fitting the corresponding quark SDE solution 
along the $p^2$ real axis or, alternatively, on a  parabola in the complex plane that does not enclose the poles. We use two pairs of complex conjugate poles as these are enough to provide a sufficiently precise fit for the quark propagator.

\bibliographystyle{unsrt}
\bibliography{main}

\end{document}